%Paper: hep-th/9303009
%From: bantay@hal9000.elte.hu (Peter Bantay)
%Date: Tue, 2 Mar 93 15:41:41

\magnification=\magstep1
\hsize=15.3 truecm
\vsize=21.0 truecm
\baselineskip=20 pt

\def\A{{\cal A}(G,\phi)}

\def\G{\vert G\vert}
\def\i{^{-1}}
\def\S{\Sigma_{g,n}}
\def\h{{\cal H}_{g,n}}
\def\mcg{\Gamma_{g,n}}

\def\c{{\cal C}_{g,n}}
\def\k{{\cal K}_{g,n}\left({x\atop y}\thinspace{a\atop b}\right)}
\def\om{\omega_z\left({ a\atop b}\right)}
\def\ze{\zeta_z\left({ x\atop y}\right)}
\def\ps{\Psi\left({ x\atop y}\thinspace{ a\atop b}\right)}
\def\xy{{x_i^{y_i}}}
\def\vp{{\cal V}_g(p_1,\dots ,p_n)}
\def\Ps{\Upsilon_{g,n}\left({ x\atop y}\right)}

\def\hp{{\cal H}_g(p_1,\dots ,p_n)}

\def\aa{a_j}
\def\ab{a_{j+1}}
\def\ba{b_j}
\def\bb{b_{j+1}}

\def\z{Z\left({ x\atop y}\thinspace{ a\atop b}\right)}
\def\o{{\cal O}[X]}
\def\cor{\langle{\cal O}_1(P_1)\dots{\cal O}_n(P_n)\rangle_\Sigma}
\def\fu{\prod_{i=1}^n{\cal O}_i[X(P_i)]}
\def\Sig{\Sigma\backslash\{P_1,\dots ,P_n\}}

\footline={\hss\tenrm\folio\hss}
\nopagenumbers
\null
\line{ISSN 0133-462X\hfil{ITP Budapest Report No. 499}}

\line{\hfil{February 1993}}

\vskip 4 truecm
\centerline{Algebraic Aspects of Orbifold Models}
\vskip 2 truecm
\centerline{P. B\'antay}
\bigskip
\centerline{Institute for Theoretical Physics}
\centerline{E\"otv\"os University, Budapest}
\vskip 4 truecm
\underbar{Abstract} : Algebraic properties of orbifold models on arbitrary
Riemann surfaces are investigated.
The action of mapping class group transformations and of standard geometric
operations is given explicitly.
An infinite dimensional extension of
the quantum group is presented.
\vfill\eject

\pageno=1
\centerline{\bf 1. Introduction}
\bigskip\bigskip

The interest in studying orbifold models is twofold. First, they provide a mean
to construct new conformal field theories from known ones,
whose properties may in principle
be determined from the knowledge of the original theory [1]. This may prove
important
in attempts to classify CFTs. Secondly, they provide an interesting laboratory
for understanding the basic properties of CFTs as well as their relation
to 3-dimensional topological field theories ( via Dijkgraaf-Witten theory [2] )
and quantum groups [3,4]. An important property of orbifold models based on
finite
groups is that the associated quantum group is finite dimensional and
semisimple,
which greatly simplifies the analysis of their algebraic properties.

The aim of the present paper is to study the algebraic properties of
the correlation functions of orbifold models on higher genus Riemann surfaces.
As we shall see, a lot of information may be obtained simply by considering
the symmetry properties of the model.

The basic property of orbifold models is that their Hilbert-space affords a
representation of the orbifold quantum group. For holomorphic orbifolds the
irreps
of the chiral algebra are in one-to-one correspondence with the irreps of the
quantum
group, and the fusion rules of the chiral algebra as well as the modular
transformations
of the genus one characters are determined by the corresponding properties of
the
quantum group [3,4]. As far as representation theory is concerned, one can
replace the
chiral algebra by the quantum group.

Our starting point is the generalized version of Verlinde's
formula [5], which gives the dimension of the space $\vp$ of holomorphic
blocks of genus $g$ with external legs saturated by the irreps
$p_1,\dots ,p_n$ of the chiral algebra $\cal A$ in terms of the matrix
elements of the modular transformation $S$ :
$$\dim\vp=\sum_q S_{0q}^{2(1-g)}\prod_{i=1}^n{S_{p_i q}\over
S_{0q}}\eqno(1.1)$$
where the sum runs over the irreps of the chiral algebra and $0$
labels the vacuum representation ( which contains the identity
operator ).

We shall interpret the rhs. of $(1.1)$ as the multiplicity of the
irrep $(p_1,\dots ,p_n)$ in some representation of the algebra
${\cal A}^{\otimes n}$ ( clearly the irreps of the
algebra ${\cal A}^{\otimes n}$ are in one-to-one correspondence
with $n$-tuples of irreps of $\cal A$ ).

Thus, the basic point of our approach is to associate to each (
topological equivalence class of ) Riemann surface $\S$ of genus
$g$ with $n$ punctures, a linear space $\h$ affording a
representation of the algebra ${\cal A}^{\otimes n}$,
subject to the restriction that the multiplicity of the irrep
$(p_1,\dots ,p_n)$ of ${\cal A}^{\otimes n}$ is equal to $\dim\vp$.

This is somewhat reminescent of topological field theories, but
there is a major difference : TFTs associate to a Riemann
surface $\S$ with punctures marked by the irreps $p_1,\dots ,p_n$
the finite dimensional vector spaces $\vp$ of holomorphic blocks [6],
while in our approach the punctures are {\it not marked}.

We shall show that the spaces $\h$ afford unitary
representations of the mapping class groups $\mcg$ of Riemann
surfaces of genus $g$ with $n$ coordinatized punctures.
Moreover, we shall introduce on the spaces $\h$ algebraic
operations which correspond to the standard geometric operations
on Riemann surfaces [7], and shall show how
the direct sum of the spaces $\h$ becomes a bigraded associative algebra
whose zero mode subalgebra may be naturally identified with the
quantum group $\A$.

Finally, we shall give the physical interpretation of the above results.
A short appendix is also included to recall the basic facts about
the orbifold quantum group and to fix the notation.

\vfill\eject

\centerline{\bf 2. The spaces $\h$}
\bigskip\bigskip

In this section we shall construct for holomorphic orbifold models the spaces
$\h$ alluded to in the introduction.

Let $G$ be a finite group and $\phi$ a unimodular 3-cocycle of $G$, and
let's denote by $\A$ the corresponding orbifold quantum group (
see the appendix for notations and definitions ).

Let $\c$ be the space of complex valued functions on $G^{2(n+g)}$.  We shall
denote the elements of $\c$ by $\ps$, where $x$ and $y$ are $n$-tuples with
components $x_i, y_i\in G\quad (i=1,\dots ,n)$, while $a$ and $b$ are
$g$-tuples with components $a_j, b_j\in G\quad (j=1,\dots ,g)$.  The elements
$z\in G$ operate on these arrays componentwise, so for example $zy$ and $x^z$
have components $zy_i$ and $x_i^z\quad (i=1,\dots ,n)$ respectively.  We shall
also use the notations $X_i=\prod_{k=1}^ix_k\quad(i=1,\dots ,n)$ and
$A_j=\prod_{k=1}^j[a_k,b_k]\quad (j=1,\dots ,g)$.

We define operators $P_i(z)$ and $Q_i(z)\quad (i=0,\dots ,n)$ acting on
$\c$. First,
$$P_0(z)\ps=\delta_{z,X_nA_g}\ps\eqno(2.1)$$
$$Q_0(z)\ps=\eta_z(X_n,A_g)\ze\om\Psi\left({ x^z\atop z\i y}\thinspace
             { a^z\atop b^z}\right)\eqno(2.2)$$
where
$$\ze=\prod_{i=1}^n\eta_z(X_{i-1},x_i)\theta_{\xy}(y_i\i ,z)\eqno(2.3)$$
and
$$\om=\prod_{j=1}^g{{\eta_z(A_{j-1},[a_j,b_j])\eta_z(a_j\i,a_j^{b_j})
\eta_z(a_j,b_j)}\over{\eta_z(a_j\i ,a_j)\eta_z(b_j,a_j^{b_j})}}\eqno(2.4)$$
\vfill\eject

Next, for $i=1,\dots ,n$
$$P_i(z)\ps=\delta_{z\i,\xy}\ps\eqno(2.5)$$
and
$$Q_i(z)\ps=\theta\i_{\xy}(z,z\i y_i\i)
\eta\i_z(x_i^{y_i},(x_i\i)^{y_i})
\Psi\left({\dots\atop\dots}{ x_i\atop  y_iz}{\dots\atop\dots}{ a\atop
b}\right)\eqno(2.6)$$
the last notation meaning that only the $i$-th argument of $\Psi$
has been changed.

Let  ${\cal A}_i$ denote the algebra generated by the
operators $P_i(z)$ and $Q_i(z)$ \break $(i=0,\dots ,n)$.
It is staightforward to show that each ${\cal A}_i$ is isomorphic to
the algebra $\A$, and they commute with each other.

We define $\h$ to be the subspace of $\c$ left invariant by ${\cal A}_0$.
$$\eqalign{\Psi\in\h\quad\Leftrightarrow\quad  \ps &=0\quad {\rm if}
\quad X_nA_g\ne 1 \quad{\rm and}\quad\cr
\ps &=\eta_z(X_n,A_g)\ze\om\Psi\left({ x^z\atop z\i y}\thinspace
             { a^z\atop b^z}\right)\cr}\eqno(2.7)$$

Clearly, $\h$ is the image of the projection operator ${\cal P}_0$, where
$${\cal P}_i={1\over\G}\sum_{z\in G}P_i(1)Q_i(z)\eqno{(i=0,\dots ,n)\qquad
(8)}$$

It follows from what have been said above that $\h$ affords a representation of
$\otimes_{i=1}^n {\cal A}_i$. Let us denote the character of this
representation by $\Ps$, which is by definition the trace over $\h$ of
the operator $\prod_{i=1}^n P_i(x_i)Q_i(y_i)$. It is given by the following
formula

$$\Ps=\sum_p S_{0p}^{2-2g-n}\prod_{i=1}^n \bar\psi_p(y_i,x_i)\eqno(2.9)$$

If we label the irreducible representations of the algebra
$\otimes_{i=1}^n {\cal A}_i$
by the $n$-tuples $(p_1,\dots ,p_n)$, where the $p_i$-s are
irreps of $\A$, then the multiplicity of the irrep $(p_1,\dots ,p_n)$ in $\h$
is
$$\sum_qS_{0q}^{2(1-g)}\prod_{i=1}^n{S_{p_iq}\over{S_{0q}}}\eqno(2.10)$$
and this last expression is just the rhs. of $(1.1)$, thus $\h$ is
indeed the space we have been looking for.

For later use, we also introduce the projection operators
$${\cal P}_{(p_1,\dots ,p_n)}=\prod_{i=1}^n S_{0p_i}\sum_{x_i,y_i\in G}
\bar\psi_{p_i}(x_i,y_i)P_i(x_i)Q_i(y_i)\eqno(2.11)$$
which project $\h$ onto the subspace $\hp$ affording the irrep $p_i$ at the
$i$-th puncture.

For $n>0$ define
$$\k=\delta_{1,X_nA_g}
{{\theta_{X_n}(y_n,y_n\i)\eta_{y_n\i}((X_n\i)^{y_n},X_n^{y_n})}
\over\phi(X_n\i,X_{n-1},x_n)}$$
$$\prod_{i=1}^n{\phi(X_i\i,X_{i-1},x_i)\over\phi(x_i\i,X_{i-1}\i,X_{i-1})}
{\eta_{y_n\i}((x_i\i)^{y_n},(X_{i-1}\i)^{y_n})\over
{\theta_{(x_i\i)^{y_n}}(y_n\i
y_i,y_i\i)\eta_{y_i\i}((x_i\i)^{y_i},x_i^{y_i})}}$$
$$\prod_{j=1}^g{{\eta_{y_n}(b_j,(a_j\i)^{b_j})\eta_{y_n}(a_j,a_j\i)}
\over{\eta_{y_n}([b_j,a_j],A_{j-1}\i)\eta_{y_n}(a_j\i,b_j)
\eta_{y_n}((a_j\i)^{b_j},a_j)}}$$
$$\prod_{j=1}^g{{\phi(A_{j-1}\i,A_{j-1},[a_j,b_j])\phi(a_j,a_j\i,a_j^{b_j})}
\over{\phi([b_j,a_j],A_{j-1}\i,A_j)\phi((a_j\i)^{b_j},a_j,[a_j,b_j])
\eta_{b_j}(a_j\i,a_j)}}\eqno(2.12)$$
\bigskip

It may be shown that $\k$ belongs to $\h$, that is
$$\k=\eta_z(X_n,A_g)
\ze\om{\cal K}_{g,n}\left({x^z\atop z\i y}{a^z\atop b^z}\right)\eqno(2.13)$$

It follows that for $n>0$ any element $\Psi$ of $\h$ has the form
$$\ps=\k\Psi^\circ\left({x_2^{y_1}\atop y_1\i y_2}{\dots\atop\dots}
{x_n^{y_1}\atop y_1\i y_n}\thinspace{a_1^{y_1}\atop b_1^{y_1}}
{\dots\atop\dots}{a_g^{y_1}\atop b_g^{y_1}}\right)\eqno(2.14)$$
where the function $\Psi^\circ$ belongs to ${\cal C}_{g,n-1}$.

There is a  natural scalar product  on $\h$
with respect to which the operators $Q_i(z)$ are unitary and
the $P_i(z)$ are hermitian. It is given by

$$\langle \Psi_1,\Psi_2\rangle ={1\over\G^{(n+g)}}\sum_{x,y,a,b}
\Psi_1\left({x\atop y}\thinspace{a\atop b}\right)
\bar\Psi_2\left({x\atop y}\thinspace{a\atop b}\right)\eqno(2.15)$$
\vfill\eject

\centerline{\bf 3. Mapping class group representations}
\bigskip\bigskip

The space $\h$ affords a unitary representation of the mapping
class group $\mcg$.
We shall illustrate the above point for the
groups $\Gamma_{0,n}$ and $\Gamma_{g,0}$.

$\Gamma_{0,n}$ is generated by the Dehn-twists
$\tau_i\quad (i=1,\dots, n)$ around a curve encircling the
$i$-th puncture and the Dehn-twists $\sigma_i$
$(i=1,\dots ,n-1)$ around a curve encircling the $i$-th and
$i+1$-th punctures.  The action of these operators on
${\cal H}_{0,n}$ is given by the following formulas

$$\tau_i\Psi\left({ x\atop y}\right)=\theta\i_{\xy}(y_i\i x_i\i,x_i)
\Psi\left({\dots\atop\dots}{ x_i\atop
x_iy_i}{\dots\atop\dots}\right)\eqno(3.1)$$
$$\sigma_i\Psi\left({ x\atop y}\right)={\phi(X_{i-1},x_{i+1},x_i^{x_{i+1}})
\over\phi(X_{i-1},x_i,x_{i+1})}\theta_{\xy}(y_i\i,x_{i+1})
\Psi\left({\dots\atop\dots}{ x_{i+1}\atop y_{i+1}}
\thinspace{ x_i^{x_{i+1}}\atop x_{i+1}\i
y_i}{\dots\atop\dots}\right)\eqno(3.2)$$
\bigskip

They satisfy the  relations
$$\sigma_i\sigma_j=\sigma_j\sigma_i\quad{\rm if}\quad \vert i-j\vert
>1\eqno(3.3)$$
$$\sigma_i\sigma_{i+1}\sigma_i=\sigma_{i+1}\sigma_i\sigma_{i+1}\eqno(3.4)$$
$$\tau_i\tau_j=\tau_j\tau_i\eqno(3.5)$$
$$\sigma_i\tau_i\sigma_i\i =\sigma_i\i\tau_i\sigma_i=\tau_{i+1}\eqno(3.6)$$
$$\sigma_1\dots\sigma_{n-1}^2\dots\sigma_1=\tau_1^2\eqno(3.7)$$
$$\left(\sigma_1\dots\sigma_{n-1}\right)^n=\tau_1\dots \tau_n\eqno(3.8)$$
\bigskip
which are the defining relations of the group $\Gamma_{0,n}$ [5].
\vfill\eject

Moreover, the $\tau_i$-s commute with the ${\cal A}_j$-s, as
well as the $\sigma_i$-s if $j\ne i,i+1$, while
$$\sigma_iP_i(z)\sigma_i\i =\sigma_i\i P_i(z)\sigma_i=P_{i+1}(z)\eqno(3.9)$$
$$\sigma_iQ_i(z)\sigma_i\i =\sigma_i\i Q_i(z)\sigma_i=Q_{i+1}(z)\eqno(3.10)$$
Consequently, $\sigma_i$ maps the space ${\cal H}_g(\dots ,p_i,p_{i+1},\dots)$
onto ${\cal H}_g(\dots ,p_{i+1},p_i,\dots)$, and this shows clearly its
braiding nature. We also have

$$\tau_i{\cal P}_{(p_1,\dots ,p_n)}=\lambda_{p_i}{\cal P}_{(p_1,\dots
,p_n)}\eqno(3.11)$$
where $\lambda_p$ is the eigenvalue of the modular $T$
operator corresponding to the irreducible character $\psi_p$ (
c.f. Eq.(A.30) ).  This means that $\tau_i$
acts on $\hp$ by multiplication by $\lambda_{p_i}$.

Let's now consider the group $\Gamma_{g,0}$.
It is generated by the Dehn-twists $\alpha_j, \beta_j$ around the curves
of a canonical homology basis $a_j, b_j\quad (j=1,\dots ,g)$ and the Dehn-twits
$\gamma_j$ around the curves $c_j\quad (j=1,\dots ,g-1)$ connecting two
consecutive handles. The action of these operators on ${\cal H}_{g,0}$ is given
by
$$\alpha_j\Psi\left({a\atop b}\right)=\theta_{a_j}(a_j,b_j)\Psi\left(
{\dots\atop\dots}{a_j\atop a_jb_j}{\dots\atop\dots}\right)\eqno(3.12)$$

$$\beta_j\Psi\left({a\atop b}\right)={\phi(a_j\i,b_j,b_j\i a_j^{b_j})\over
{\phi(a_j\i,b_j,b_j\i a_j)\phi(b_j,b_j\i a_j,b_j)}}\Psi\left(
{\dots\atop\dots}{b_j\i a_j\atop b_j}{\dots\atop\dots}\right)\eqno(3.13)$$

$$\gamma_j\Psi\left({a\atop b}\right)=\varsigma
\Psi\left({\dots\atop\dots}
{c\i a_j\atop b_j^c}\thinspace{a_{j+1}c\atop b_{j+1}}
{\dots\atop\dots}\right)\eqno(3.14)$$
where

$$c=\ba(\bb\i)^{\ab}$$
and the prefactor is

$$\varsigma={\phi(A_{j-1},[\aa,\ba][\ba,c],[c,\ba][\ab,\bb])\over
\phi(A_{j-1},[\aa,\ba],[\ab,\bb])}
{\eta_{\ba}(c\i,c)\over\eta_{\bb}(\ab,c)}$$
$${{\phi(\ba,\aa^{\ba},[\ba,c])\phi([\aa,\ba],[\ba,c],[c,\ba][\ab,\bb])
\phi(\aa\i,c,(\aa c\i)^{\ba c})}\over
{\phi(\aa,\ba,[\ba,c])\phi([\ba,c],[c,\ba],[\ab,\bb])
\phi(\aa\i,\aa^{\ba},[\ba,c])}}$$
$${{\phi(c\i,\ab\i,(\ab c)^{\bb})\phi(\aa\i c,c\i,\aa)\phi(c\i\ab\i,\ab,c)}
\over{\phi(\ab\i,\ab^{\bb},c^{\bb})\phi(\ba,c,(\aa c\i)^{\ba c})
\phi(c\i,\ab\i,\ab)}}$$
$${{\phi(c\i,\aa,\ba^c)\phi((c\i)^{\ba},c,c\i)\phi(\ba,(c\i)^{\ba},c)}\over
{\phi(c\i,\ba c,(\aa c\i)^{\ba c})\phi(\aa\i,c,c\i)\phi(c\i,\ba,c)}}$$
$${{\phi(c\i,[\ab,\bb],\bb\i\ba)\phi(\ba\i,c\bb,\bb\i\ba)
\phi(\bb\i\ba,\ba\i,c\bb)}\over{\phi([\ba,c],c\i,c^{\ba})
\phi(c\i [\ab,\bb],\bb\i\ba,[\ab,\bb])\phi(\bb,\bb\i,\ba)}}$$
$${{\phi(c,\bb,\bb\i \ba)\phi(\bb\i,\bb,c^{\bb})\phi(\bb,\bb\i,\bb)}\over
{\phi(c\i,[\ab,\bb],c^{\bb})\phi(\ba\i,\ba,c^{\ba})\phi(\bb\i,\ba,\ba\i)
\phi(\ba,\ba\i,\ba)}}\eqno(3.15)$$
\bigskip

It may be shown by rather lengthy calculations, that the above operators
satisfy
the defining relations of the group $\Gamma_{g,0}$ ( cf. [5] ).

\vfill\eject

\centerline{\bf 4. Geometric operations}
\bigskip\bigskip

Until now we have only considered operators mapping $\h$ to itself. We now turn
to the study of operators between different spaces.

The first operators of interest are the projections
${\cal P}_i$ defined by $(2.8)$, which correspond to
saturating the $i$-th puncture by the vacuum.
The image of $\h$ under this projection may be naturally identified with the
space ${\cal H}_{g,n-1}$. This provides us with a natural imbedding of
${\cal H}_{g,m}$ into $\h$ for $m<n$.

For $\Psi_1 \in {\cal H}_{g,n+1}$ and
$\Psi_2\in {\cal H}_{k,m+1}$ we define $\Psi_1\infty\Psi_2\in
{\cal H}_{g+k,n+m}$ as
$$\Psi_1\infty\Psi_2\left({x_1\atop y_1}{\dots\atop\dots}{x_{n+m}\atop y_{n+m}}
\thinspace{a_1\atop b_1}{\dots\atop\dots}{a_{g+k}\atop b_{g+k}}\right)=$$
$$\sum_{X,Y}\xi_{X,Y}\Psi_1\left({x_1\atop y_1}
{\dots\atop\dots}{x_n\atop y_n}\thinspace{X\atop Y}\thinspace
{a_{k+1}\atop b_{k+1}}{\dots\atop\dots}{a_{g+k}\atop b_{g+k}}\right)
\Psi_2\left({X\i\atop Y}\thinspace{x_{n+1}\atop y_{n+1}}
{\dots\atop\dots}{x_{n+m}\atop y_{n+m}}\thinspace{a_1\atop b_1}
{\dots\atop\dots}{a_k\atop b_k}\right)\eqno(4.1)$$
where
$$\xi_{X,Y}=\eta_{Y\i}((X\i)^Y,X^Y)
{{\phi(A_{k+g}\i A_k,A_k\i A_{k+g},A_{k+g}\i A_k)
\phi(A_k\i A_{k+g},A_{k+g}\i,A_{k+g})}\over\phi(X\i X_n\i, X_n, X)}$$
$$\prod_{i=1}^m\phi(X\i X_n\i,X_{n+i-1},x_{n+i})
\prod_{j=1}^g\phi(A_k\i,A_{k+j-1},[a_{k+j},b_{k+j}])
\eqno(4.2)$$
\medskip

This operation, which may be shown to be associative,
corresponds to sewing two  Riemann surfaces.
Endowed with this operation, the space
$${\cal H}=\bigoplus_{n>0,g\ge 0}\h\eqno(4.3)$$
becomes a bigraded associative algebra.

Clearly the subspace ${\cal H}_{0,2}$ is a closed subalgebra of
$\cal H$ ( it is the zero grade subalgebra ). This subalgebra is
isomorphic to $\A$.
This may be shown by introducing the following basis in ${\cal H}_{0,2}$

$$\vert X,Y\rangle\left({x_1 \atop y_1}\thinspace{x_2\atop y_2}\right)=
{\theta_X(Y,y_2\i)\over \eta_{y_1\i}(X\i,X)}
\delta_{x_1x_2,1}\delta_{X\i,x_1^{y_1}}\delta_{Y,y_1\i y_2}\eqno(4.4)$$
\medskip
It is then straightforward to show that

$$\vert x_1,y_1\rangle\infty\vert x_2,y_2\rangle=\delta_{x_2,x_1^{y_1}}
\theta_{x_1}(y_1,y_2)\vert x_1,y_1y_2\rangle\eqno(4.5)$$

For $\Psi$ in $\cal H$ we have
$$\vert x,y\rangle\infty\Psi=P_1(x)Q_1(y)\Psi\eqno(4.6)$$
This shows that the action of ${\cal A}_1$ on $\cal H$ may be
realized by sewing with elements of ${\cal H}_{0,2}$. Recalling Eqs.(3.9)
and (3.10),we see that the action of the other ${\cal A}_i$-s may
be realized as well by combining $\infty$ with suitable braids (
which corresponds to sewing in different channels ).

Of special interest is the infinite dimensional subalgebra
$$\hat\A=\bigoplus_{g\ge 0}{\cal H}_{g,2}\eqno(4.7)$$
which may be viewed as the extension of the original algebra
$\A$ to all genera.

Another operation of interest is the map ${\cal S}:{\cal H}_{g,n+2}\to {\cal
H}_{g+1,n}$
defined by
$${\cal S}\Psi\left({x\atop y}\thinspace{a_0\atop b_0}\thinspace{a\atop
b}\right)=
{{\phi(X_n,[a_0,b_0],A_g)\phi(X_n,a_0\i ,a_0^{b_0})}\over{
\prod_{j=1}^g\phi([a_0,b_0],A_{j-1},[a_j,b_j])}}$$
$$\sum_{z\in G}\eta_{z\i}((a_0\i)^z,a_0^z)\theta_{a_0^z}(z\i,b_0)
\Psi\left({x\atop y}\thinspace{a_0\i\atop z}\thinspace
{a_0^{b_0}\atop b_0\i z}\thinspace{a\atop b}\right)\eqno(4.8)$$

This operation corresponds to sewing the $n$-th and $n+1$-th
punctures of the same Riemann surface, and it maps $\h$
onto ${\cal H}_{g+1,n-2}$.

For example, $\cal S$ maps ${\cal H}_{0,2}$ onto ${\cal H}_{1,0}$.
The matrix elements of $\cal S$ between the basis $\vert x,y\rangle$ of
${\cal H}_{0,2}$ ( c.f. Eq.(4.4) )
and the natural basis of ${\cal H}_{1,0}$  provided by the
irreducible characters $\psi_p$ of $\A$ are

$$\langle\psi_p\vert{\cal S}\vert x,y\rangle=\psi_p(x,y) \eqno(4.9)$$

Next, for $i=1,\dots ,n$ consider the maps
$\partial_i:{\cal H}_{g,n+1}\to {\cal H}_{g,n}$ given by
$$\partial_i\Psi\left({x\atop y}\thinspace{a\atop b}\right)=$$
$$\sum_{z\in G}
{\phi(X_{i-1},z,z\i x_i)\over\phi(X_{i-1}^{y_i},z^{y_i},(z\i x_i)^{y_i})}
\eta_{y_i\i}(z^{y_i},(z\i x_i)^{y_i})\Psi\left({\dots\atop\dots}
{z\atop y_i}\thinspace{z\i x_i\atop y_i}{\dots\atop\dots}
{a\atop b}\right)\eqno(4.10)$$

They correspond to coalescing the $i$-th and $i+1$-th punctures,
and satisfy the relations
$$\partial_i\partial_j=\partial_j\partial_{i+1}\qquad{\rm for}
\qquad n > i\ge j\ge 1\eqno(4.11)$$

The above relations imply that the operator ${\cal D}_n$ defined as
$${\cal D}_n=\sum_{i=1}^n(-1)^i\partial_i\eqno(4.12)$$
is a differential operator in the sense that
$${\cal D}_{n-1}{\cal D}_n=0\eqno(4.13)$$

\centerline{\bf 5. Physical interpretation }
\bigskip\bigskip

Let's consider a conformal field theory on a compact Riemann surface $\Sigma$
of genus $g$, defined via an action
functional $S[X]$. The correlation functions of local operators
( i.e. local functionals $\o$ of the $X$-s ) are given by the
functional integral
$$\cor=\int{\cal D}X\fu\exp(-iS[X])\eqno(5.1)$$
where one should integrate over those $X$-s which are
single-valued on $\Sigma$.
Besides the insertion points $P_i\in\Sigma$, the correlator (5.1) depends also
on the moduli of the Riemann surface $\Sigma$.

To get an orbifold model [8], we choose a group $G$ of global
symmetries , i.e. transformations of the $X$-s that
leave the action invariant.  For simplicity, we shall assume
that $G$ is finite. The orbifold model is defined
by imposing equivalence modulo $G$, i.e. gauging the group $G$.
The correlators of the orbifold model are still given by the functional
integral (5.1), but with integration to be performed over
$X$-s which are allowed to have $G$ valued monodromies
around the homotopicaly nontrivial curves of the punctured surface
$\Sig\equiv\S$.

The multivaluedness of $X$ is completely characterized by its
monodromies around the simple closed curves on $\S$.  It is
enough to specify the monodromies around a set of generators of
the fundamental group of $\S$. A convenient set of generators
of the fundamental group is obtained by fixing a base point
$P_0$ on $\Sig$, and considering the simple loops based at $P_0$
which encircle one puncture $P_i$ at a time, as well as the
curves of a canonical homology basis of the compact Riemann
surface  $\Sigma$. We shall denote the monodromies around these
loops by $x_1,\dots ,x_n$ and $a_1,\dots ,a_g,b_1,\dots ,b_g$
respectively, and call their collection a twist structure.
They have to satisfy
$$\prod_{i=1}^n x_i\prod_{j=1}^g [a_j,b_j]=1\eqno(5.2)$$
in order that there
exists an $X$ with these given monodromies.

The functional integral for the
orbifold model decomposes into a sum
$$\cor={1\over\G^{(n+g)}}\sum_{x,y,a,b}\z\eqno(5.3)$$
where $x,y$ and $a,b$ are respectively $n$ and $g$-tuples of
elements of $G$, while
$$\z=\int{\cal D}X\prod_{i=1}^n{\cal O}_i[y_i X(P_i)]\exp(-iS[X])\eqno(5.4)$$
where integration is restricted to those $X$-s with monodromies
specified by $x,a$ and $b$. The sum over $y_i$-s in (5.3)
implements the $G$-invariant projection and it is the analogue of
the GSO projection in fermionic string theory, while the sum over the
twist structures is the analogue of the sum over spin
structures.

The basic properties of $\z$ are
$$\z=0\qquad{\rm if}\qquad \prod_{i=1}^n x_i\prod_{j=1}^g [a_j,b_j]\ne
1\eqno(5.5)$$
which follows from (5.2), and
$$\z=Z\left({ x^z\atop z\i y}\thinspace{ a^z\atop b^z}\right)\eqno(5.6)$$
which follows from the $G$-invariance of the action.

Let us suppose now that we are considering a holomorphic
orbifold model, i.e. the theory we started with had only one chiral
block. Then  $\z$ is the modulus squared of some
holomorphic function of the moduli, i.e.
$$\z=\Bigl\vert\ps\Bigr\vert^2\eqno(5.7)$$
where $\ps$ is a holomorphic function of the moduli of
$\Sig$. The point is that in order to have a consistent theory,
$\Psi$ should belong to $\h$. This claim follows from the observation that
(5.5)
implies a similar condition on $\Psi$, while (5.6) implies that
$$\ps=\vartheta\Psi\left({x^z \atop z\i y}\thinspace{a^z\atop
b^z}\right)\eqno(5.8)$$
where the phase $\vartheta$ depends only on the variables $x,y,a,b$, but
not on the operators ${\cal O}_i$ nor the moduli of the surface. Duality and
modular invariance of the theory then forces $\Psi$ to belong to $\h$.

If we specify the irreps $p_1,\dots ,p_n$ of the chiral algebra to which the
operators ${\cal O}_1,\dots ,
{\cal O}_n$ belong, then $\Psi$ may be considered as an intertwiner between the
irrep
$(p_1,\dots ,p_n)$ of ${\cal A}^{\otimes n}$ and the space $\h$, which depends
holomorphicaly on the moduli. An argument involving Schur's lemma shows that
$\Psi$
will depend on $\dim\vp$ holomorphic functions of the moduli, and consequently
the correlator (5.3) will be the sum of the modulus squared of these functions,
which
of course are nothing but the holomorphic blocks of the orbifold model.

If we consider $\Psi$ as a function over Teichmuller space ${\cal T}_{g,n}$,
then it
should be equivariant with respect to $\mcg$, i.e. the action of any mapping
class group
transformation on ${\cal T}_{g,n}$ can be compensated by the action of its
inverse on $\h$.

The physical meaning of the operations introduced in the previous sections is
quite
transparent in this setting. For example the operations $\infty$ and $\cal S$
are related
to the sewing of correlators, while the operations $\partial_i$ are related to
operator
product expansions. The associativity of $\infty$ is connected to the duality
of the theory, while (4.11) is connected to the associativity of OPEs.

\vfill\eject

\centerline{\bf 6. Summary}
\bigskip\bigskip

In this paper we have studied the properties of orbifold models
on higher genus Riemann surfaces. We have done this by
associating to the Riemann surface $\S$ the space $\h$, which
carries a specific representation of the $n$-fold tensor power
${\cal A}^{\otimes n}$ of the orbifold quantum group. We
have shown that the mapping class group $\mcg$ is represented
on the space $\h$, as well as other important geometrical
operations. It turned out that the algebraic structure of the
quantum group is closely related to the operation of sewing
Riemann surfaces, and this led us to generalize the quantum
group to higher genera.

We have sketched only those aspects of the theory which have
a direct physical interpretation. The spaces $\h$ admit more
structure than described in the paper, which may prove useful
in mathematical applications.

One such aplication is related to Moonshine [9,10]. Starting from a
holomorphic CFT whose symmetry group contains the group $G$
-- which is known to exist for the Monster [11] --, one may
in principle compute the functions $\Psi\in\h$ for suitable
vertex operators, and these would satisfy a set of relations
which may be naturally interpreted as an extension of the
Moonshine properties to arbitrary Riemann surfaces. From this
point of view the two basic facets of Moonshine are the existence
of a holomorphic CFT on which $G$ acts as a symmetry group and
the theory developped in section 3.

Another
interesting application of the above theory  is to the description
of the statistics of many-vortex systems in spontaneously broken gauge
theories, which will be
dealt with in a forthcoming publication [12].

\vfill\eject

\null

\centerline{\bf Appendix}
\bigskip\bigskip

Let $G$ be a finite group, and $\phi$ a unimodular 3-cocycle of $G$,
that is a  function on $G^3$ taking its values in the complex unit
circle, and which satisfies
$$\phi(x_1,x_2,x_3)\phi(x_1,x_2x_3,x_4)\phi(x_2,x_3,x_4)=
\phi(x_1x_2,x_3,x_4)\phi(x_1,x_2,x_3x_4)\eqno(A.1)$$
We suppose $\phi$ normalized, i.e. it takes the value 1 whenever any
of its arguments equals the identity element of $G$ ( which we
shall also denote by 1 ).

We introduce the auxiliary quantities ( where $x^z\equiv z\i xz$ )
\bigskip
$$\theta_z(x,y)={{\phi(x,z^x,y)}\over{\phi(z,x,y)\phi(x,y,z^{xy})}}\eqno(A.2)$$
$$\eta_z(x,y)={{\phi(x,z,y^z)}\over{\phi(x,y,z)\phi(z,x^z,y^z)}}\eqno(A.3)$$
\bigskip

They satisfy the following basic identities
\medskip
$$\theta_z(x_1,x_2)\theta_z(x_1x_2,x_3)=\theta_{z^{x_1}}(x_2,x_3)
\theta_z(x_1,x_2x_3)\eqno(A.4)$$
$$\theta_{z_1}(x_1,x_2)\theta_{z_2}(x_1,x_2)\eta_{x_1}(z_1,z_2)
\eta_{x_2}(z_1^{x_1},z_2^{x_1})=\theta_{z_1z_2}(x_1,x_2)\eta_{x_1x_2}(z_1,z_2)
\eqno(A.5)$$
$$\eta_z(x_1,x_2)\eta_z(x_1x_2,x_3)\phi(x_1,x_2,x_3)=\eta_z(x_2,x_3)
\eta_z(x_1,x_2x_3)\phi(x_1^z,x_2^z,x_3^z)\eqno(A.6)$$
$${\theta_x(z,y^z)\over{\theta_x(y,z)}}={\eta_z(y,x^y)\over{\eta_z(x,y)}}
\eqno(A.7)$$
$$\theta_z(x,y)=\eta_z(x,y)\qquad{\rm if}\quad x,y\in C_G(z)\eqno(A.8)$$
\vfill\eject

$\A$ is the associative algebra generated
by the elements $P(x)$ and $Q(x)$ ( $x\in G$ ), subject to the relations
$$P(x)P(y)=\cases{P(x)  & if $x=y$, \cr
                  0     & otherwise \cr}\eqno(A.9)$$
$$Q^{-1}(x)P(y)Q(x)=P(y^x)\eqno(A.10)$$
$$Q(x)Q(y)=\sum_{z\in G}\theta_z(x,y)P(z)Q(xy)\eqno(A.11)$$
$$\sum_{x\in G}P(x)=Q(1)\eqno(A.12)$$
\bigskip
$\A$ is semisimple, with unit element $Q(1)$.

The comultiplication, a homomorphism $\Delta : \A\to\A\otimes\A$,
is defined by
$$\Delta P(x)=\sum_{z\in G}P(z)\otimes P(z^{-1}x)\eqno(A.13)$$
$$\Delta Q(x)=\sum_{y,z \in G}\eta_x(y,z)P(y)Q(x)\otimes P(z)Q(x)\eqno(A.14)$$

\bigskip
The comultiplication $\Delta$ is needed in order to define the tensor
product of $\A$-modules. The following properties of $\Delta$ insure
that this tensor product is associative and commutative, or in the
language of Hopf algebras, $\A$ endowed with $\Delta$ is a
quasi-triangular quasi-Hopf algebra.
$$(id\otimes\Delta)\circ\Delta(a)=\varphi((\Delta\otimes
id)\circ\Delta(a))\varphi^{-1},\eqno(A.15)$$
$$\tau\circ\Delta(a)=R\Delta(a) R^{-1}\eqno(A.16)$$
for all $a\in\A$, where $\tau$ is the map $a\otimes b\mapsto b\otimes a$,
$$\varphi=\sum_{x,y,z\in G}\phi(x,y,z)P(x)\otimes P(y)\otimes
P(z),\eqno(A.17)$$ and
$$R=\sum_{x\in G}P(x)\otimes Q(x)\eqno(A.18)$$
is the so called universal $R$-matrix. The antipode $\gamma$ is
$$\gamma\left(P(x)Q(y)\right)=\theta_{x\i}\i(y,y\i)\eta_y\i(x\i,x)
Q(y\i)P(x\i)\eqno(A.19)$$

The representation theory of $\A$ is very similar to that of finite groups.
One defines the character $\psi$ of a representation as the trace of
$P(x)Q(y)$
$$\psi(x,y)={\rm Tr}\left(P(x)Q(y)\right)\eqno(A.20)$$

The basic properties of characters are the following :
$$\psi(x,y)=0\qquad{\rm if}\qquad xy\ne yx\eqno(A.21)$$
$$\psi(x^z,y^z)={\eta_z(y,x)\over\eta_z(x,y)}\psi(x,y)\eqno(A.22)$$
$$\psi(x,y\i)=\theta_x(y,y\i)\bar\psi(x,y)\eqno(A.23)$$
\bigskip
It may be shown that representations of $\A$ are characterized
up to equivalence by their characters. If we denote by $\psi_p$ the
characters of the irreducible
representations, we have the so-called generalized orthogonality relation
$${1\over\G}\sum_{z\in G}\theta_x\i(z,z\i y)\psi_p(x,z)\psi_q(x,z\i y)=
\delta_{p,q}{\psi_p(x,y)\over d_p}\eqno(A.24)$$
where $d_p$ is the dimension of $p$-th irrep. A direct consequence of $(A.24)$
is the orthogonality relation
$${1\over\G}\sum_{x,y\in G}\psi_p(x,y)\bar\psi_q(x,y)=\delta_{p,q}\eqno(A.25)$$
\vfill\eject
The following formulas, which follow from the orthogonality relations, are
very important in the proof of formula $(2.9)$ :

$$\sum_{{u\in C(x)}\atop {[y,u]=z}}{\eta_x(y,u)\over\eta_x(u,y^u)}=
\sum_p\psi_p(x,y)\bar\psi_p(x,yz)\eqno(A.26)$$

$${1\over\G}\sum_{{x,y\in C(z)}\atop{[x,y]=u}}
{{\eta_z(x,y)\eta_z(x\i,x^y)}\over{\eta_z(y,x^y)\eta_z(x,x\i)}}=
\sum_p d_p\i\bar\psi_p(z,u)\eqno(A.27)$$

The linear space spanned by the characters is nothing but the space
${\cal H}_{1,0}$.
Consequently the modular group $\Gamma_{1,0}\cong SL_2({\bf Z})$ is represented
unitarily on
the space of characters. It is well known that $SL_2({\bf Z})$ is generated
by the transformations $S$ and $T$ subject to the relations $S^4=1$ and
$(ST)^3=S^2$. The action of these generators on the characters is given by
$$T\psi(x,y)=\theta_x(x,y)\psi(x,xy)\eqno(A.28)$$
$$S\psi(x,y)=\theta_y\i(x,x\i)\psi(y,x\i)\eqno(A.29)$$

It may be shown that the transformation $T$ is diagonal in the basis of
irreducible characters :
$$T\psi_p(x,y)=\lambda_p\psi_p(x,y)\eqno(A.30)$$
where
$$\lambda_p={1\over d_p}\sum_{x\in G}\psi_p(x,x)\eqno(A.31)$$

Moreover, the matrix of $S$ in this same basis,
$$S_{pq}={1\over\G}\sum_{x,y\in G}\bar\psi_p(x,y)\bar\psi_q(y,x)\eqno(A.32)$$
is symmetric, $S_{pq}=S_{qp}$.

A most important result is the so-called Verlinde formula. As
explained before, the comultiplication $\Delta$ allows us to
define the tensor product of representations of the quantum
group $\A$. Let us denote by $N_{pq}^r$ the multiplicity of the
irrep $r$ in the tensor product of the irreps $p$ and $q$. The
Verlinde formula expresses $N_{pq}^r$ in terms of the matrix
elements of $S$ :
$$N_{pq}^r=\sum_s{{S_{ps}S_{qs}\bar S_{rs}}\over S_{0s}}\eqno(A.33)$$

Finally, we note that  the theory only depends on the cohomology class
of the 3-cocycle $\phi$, i.e. cohomologous cocycles give rise to isomorphic
algebras whose representation theory is identical. This "gauge invariance"
carries over to the theory developped in the main body of the paper, e.g. to
the mapping class group representations associated to the pair $(G,\phi)$.

\vfill\eject

\centerline{\bf References}
\bigskip\bigskip

\item{[1]}  R. Dijkgraaf, C. Vafa, E, Verlinde and H. Verlinde : Comm. Math.
            Phys. {\bf 123}, 485 (1989);

\item{[2]} R. Dijkgraaf and E. Witten : Comm. Math. Phys. {\bf 129}, 393
(1990);

\item{[3]} P. Bantay : Phys. Lett. {\bf B245 }, 477 (1990);
           Lett. Math.  Phys. {\bf 22}, 187 (1991);

\item{[4]} R. Dijkgraaf, V. Pasquier and P. Roche : preprint PUTP-1169 (1990);

\item{[5]} G. Moore and E. Seiberg : Comm. Math. Phys. {\bf 123}, 177 (1989);

\item{[6]}  M. Atiyah :  The Geometry and Physics of Knots, Cambridge Univ.
Press (1990);

\item{[7]} C. Vafa : Phys. Lett. {\bf B199}, 195 (1987);

\item{[8]}  L. Dixon, J. Harvey, C. Vafa and E. Witten : Nucl. Phys. {\bf
B261}, 678 (1986); Nucl. Phys. {\bf B274}, 285 (1986);

\item{[9]} J. Conway and P. Norton : Bull. London Math. Soc. {\bf 11}, 308
(1979);

\item{[10]} G. Mason : Proc. Symp. Pure Math. {\bf 47}, 208 (1987);

\item{[11]} I. Frenkel, J. Lepowsky and A. Meurman : Vertex Operator Algebras
and the Monster, Academic Press (1988);

\item{[12]} P. Bantay : in preparation;

\end